\documentclass[a4paper]{article}
\usepackage{amsmath,graphicx,hyperref,typeface,float,enumitem,multirow,booktabs,array,amssymb,tabularx}
\usepackage{caption}
\usepackage{subcaption}
\usepackage{INTERSPEECH2022}

\title{A Multi-Stage Multi-Codebook VQ-VAE Approach to \\ High-Performance Neural TTS}

\name{Haohan Guo$^*$\thanks{Work performed during the first author's internship at Xiaohongshu}, Fenglong Xie$^\ddag$, Frank K. Soong$^\dag$, Xixin Wu$^*$, Helen Meng$^*$}
\address{$^*$The Chinese University of Hong Kong, Hong Kong SAR, China \\
$^\ddag$ Xiaohongshu Inc., Shanghai, China \\
  $^\dag$Microsoft Research Asia, Beijing, China}
\email{\href{mailto:hguo@se.cuhk.edu.hk}{\nolinkurl{{hguo, wuxx, hmmeng}@se.cuhk.edu.hk}}, fenglongxie@xiaohongshu.com, frankkps@microsoft.com}

\begin{document}

\maketitle

\begin{abstract}
We propose a Multi-Stage, Multi-Codebook (MSMC) approach to high performance neural TTS synthesis. A vector-quantized, variational autoencoder (VQ-VAE) based feature analyzer is used to encode Mel spectrograms of speech training data by down-sampling progressively in multiple stages into MSMC Representations (MSMCRs) with different time resolutions, and quantizing them with multiple VQ codebooks, respectively. Multi-stage predictors are trained to map the input text sequence to MSMCRs progressively by minimizing a combined loss of the reconstruction Mean Square Error (MSE) and “triplet loss”. In synthesis, the neural vocoder converts the predicted MSMCRs into final speech waveforms. The proposed approach is trained and tested with an English TTS database of 16 hours by a female speaker. The proposed TTS achieves an MOS score of 4.41, which outperforms the baseline with an MOS of 3.62. Compact versions of the proposed TTS with much less parameters can still preserve high MOS scores. Ablation studies show that both multiple stages and multiple codebooks are effective for achieving high TTS performance.
\end{abstract}


\section{Introduction}
\label{sec:intro}

The performance of text-to-speech synthesis is affected by the deviations between ground-truth acoustic features and the predicted features \cite{ijima2016objective, saito2017statistical}. This problem is usually alleviated by enhancing the acoustic model with more advanced structures \cite{ihm2020reformer,yang2019enhancing,FastSpeech,kim2021conditional} and more complicated algorithms  \cite{li2019neural,yang21e_interspeech,glowtts}, or adapting the vocoder to the predicted features \cite{shen2018natural, Neekhara2021AdaptingTM}. However, these approaches either offer limited improvement \cite{guo2022multi}, or demand higher modeling complexity to TTS. To search for a more effective solution, we attempt to address this problem from the perspective of the feature representation. 

For TTS, a good speech representation offers high completeness, i.e. captures sufficent information for high-quality reconstruction of the target audio, which generally calls for higher dimensionality or larger space volume. However, this also increases the difficulty in modeling \cite{verleysen2005curse}. The feature with a overly large space may lead to over-fitting as inputs \cite{liu2016overfitting} or under-fitting as outputs \cite{labbe2009learning}. Hence, it is also crucial to maintain compactness of the representation with the minimal feature space. The vector-quantized variational autoencoders (VQ-VAE) offers a discrete speech representation with notable compactness in speech codecs \cite{garbacea2019low, chen2021tenc}. This inspires us to explore the use of a VQ-VAE based feature analyzer for TTS.

We first examine the vector-quantized speech representation, in terms of the trade-off for the lower completeness due to higher compactness. We then proceed to use a multi-stage multi-codebook vector-quantized representation (MSMCR), where multiple codebooks are used to quantize a vector to achieve a richer representation, and to quantize the speech sequence into multiple sub-sequences for preservation of information at different time resolutions. MSMC-TTS is proposed to generate speech from the text via MSMCR. After obtaining MSMCRs from Mel spectrograms via the feature analyzer, a neural vocoder is used to up-sample them to the audio. Meanwhile, the multi-stage predictor, as the acoustic model, maps the input text to MSMCRs progressively in resolution. It is trained with the combination of MSE and triplet loss to better obtain the expected codes. This study compares the proposed approach against baselines in various scenarios to show its effectiveness to TTS and the lower requirement to modeling complexity. Ablation studies validate that multiple stages and multiple codebooks are both crutial for MSMC-TTS.

\section{Multi-Stage Multi-Codebook Representation}
\label{sec:MSMCR}

In this section, we will describe the VQ-VAE based feature analyzer and the proposed multi-stage multi-codebook VQ-VAE.

\subsection{Vector-Quantized Variational AutoEncoder}

VQ-VAE aims to learn a discrete latent representation from target data using an encoder-decoder model. The encoder embeds the input sequence $x$ to the hidden sequence $h$, and then quantizes it by replacing each vector $h_{i}$ with the nearest prototype vector in the codebook $C^{M \times N}$ (composed of $M$ vectors with the dimension of $N$) according to the Euclidean distance. The decoder maps the quantized sequence $q$ to $\hat{x}$ for reconstruction. Since quantization is a non-differentiable operation, we usually update the codebook with the exponential moving average \cite{Oord2017NeuralDR}, and train the model with the loss function as follows:
\begin{align}
\label{eq:vqvae_loss}
    L_{VQ} = MSE(x, \hat{x}) + \alpha * MSE(h, sg(q))
\end{align}
where $MSE$ and $sg$ denotes the mean square error and stop-gradient operation, and $\alpha$ is a weight coefficient.

In this framework, an acoustic feature sequence can be represented by an index sequence. When a vector composed of $N$ float-point numbers (32 bits) is compressed to one integer with the range of $[0, M)$, the compression ratio $CR$ can be calculated as follows:
\begin{equation}
    CR = \frac{N}{1} * \frac{32}{\log_2(M)} 
\end{equation}
In this way, the 80-dim Mel spectrogram can be compressed $2560 / log_2(M)$ times. The high compression ratio brings higher compactness. However, higher compactness reduces the completeness due to information loss, and the preserved information may be insufficient for speech reconstruction and prediction. To address this problem, we propose the multi-stage multi-codebook representation.

\subsection{Multi-Stage Multi-Codebook VQ-VAE}

The MSMC-VQ-VAE based feature analyzer aims to encode a speech sequence stage-wise into the multi-stage multi-codebook representation (MSMCR), i.e. a set of sub-sequences at different time resolutions, and quantized by multiple codebooks respectively. This approach is composed of two parts -- multi-head vector quantization, and multi-stage autoencoder.

\subsubsection{Multi-Head Vector Quantization}

Reducing the compression ratio may give a richer representation, but may be difficult in VQ-VAE. To half the compression ratio, we need to square the codebook size, which leads to exponentially increasing data requirements, and also computation for training to sufficiently update each code. To alleviate this problem, we propose to quantize each vector with multiple codebooks, and hence reduce the compression ratio further: 

\begin{equation}
    CR = \frac{N}{1} * \frac{32}{\sum^H_{k=1}\log_2(M_k)} 
\end{equation}
where $H$ denotes the number of codebooks, $M_k$ denotes the number of codes in the $k$-th codebook. To obtain this multi-codebook representation, we propose multi-head vector quantization (MHVQ), as inspired by multi-head attention mechanism \cite{vaswani2017attention}. In this approach, one VQ codebook $C^{M \times N}$ is splitted into $H$ codebooks $\{C^{M \times \frac{N}{H}}_1, ..., C^{M \times \frac{N}{H}}_H\}$. The hidden vector $h_i$ is segmented into $H$ sub-vectors with the dimension of $\frac{N}{H}$, and quantized by their corresponding codebooks, finally concatenated as the output $q_i$. In this way, without introducing more paramters, the compression ratio can be halved by doubling the number of codebooks.

\subsubsection{Multi-Stage AutoEncoder}
\label{ssec:hvqr}

In addition to improving completeness, it is also crutial to well-represent all speech attributes with different time resolutions, such as low-resolutional prosody and high-resolutional pronunciation. Otherwise, any overly compressed attribute may lead to serious degradation in overall quality, as the ``Cask Effect'' reveals. In this work, we propose the multi-stage autoencoder to stagely model the speech sequence at different resolutions \cite{razavi2019generating}.

\begin{figure}[htp]
\centering
\includegraphics[width=7.5cm]{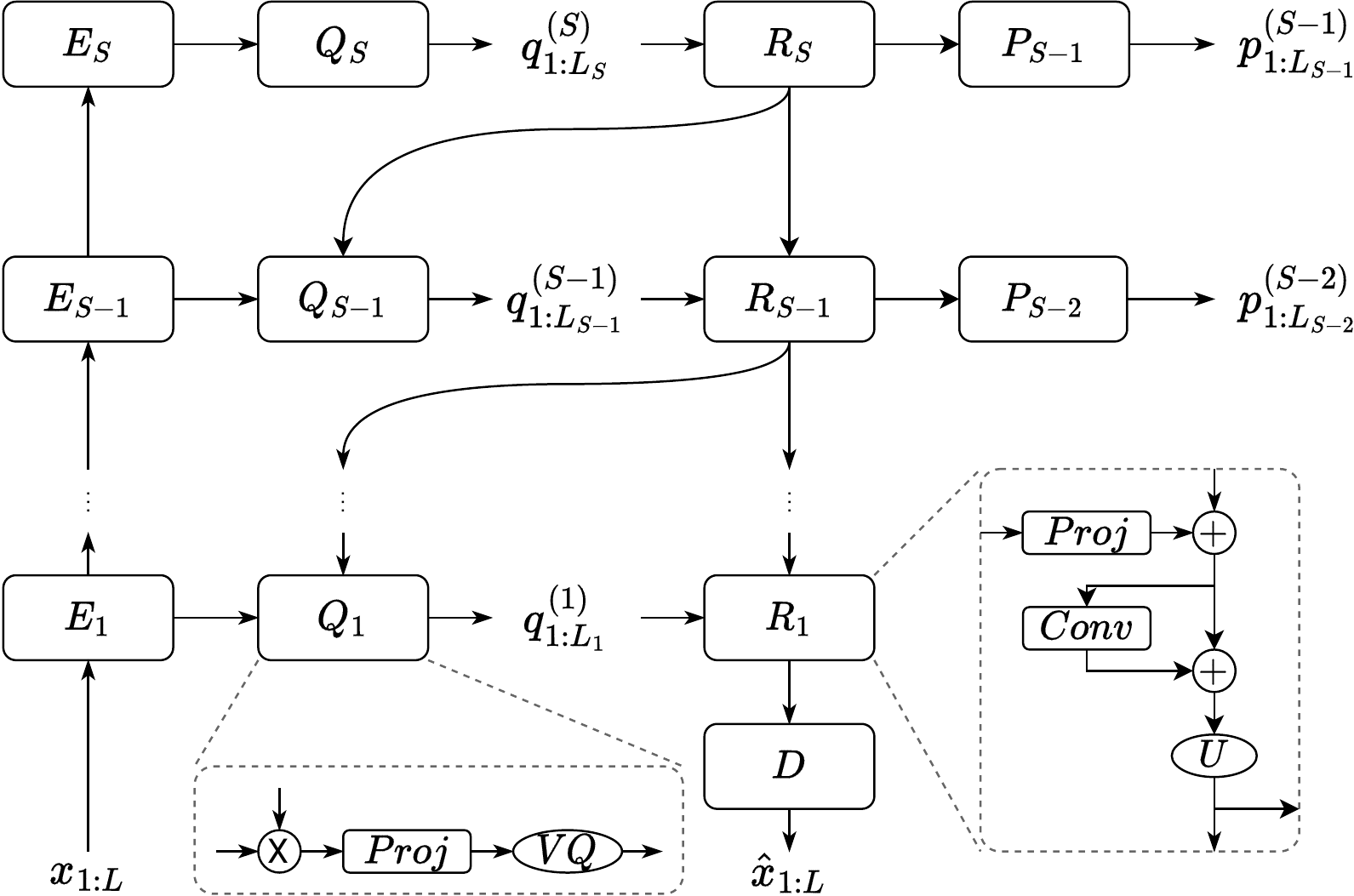}
\caption{The architecture of the MSMC-VQ-VAE. $E_S$, $Q_S$, $R_S$, $P_S$ denote the $S$-th encoder, quantizer, residual module, prediction module. $D$, $Proj$, $Conv$ denote the decoder, the projection and convolutional layer. $U$, $VQ$, ``$\times$'', ``$+$'' are upsampling, vector quantization , concatenating, and adding operations.}
\label{fig:mssr}
\end{figure}

As shown in Fig.\ref{fig:mssr}, this model aims to encode the input sequence $x_{1:L}$ into the multi-stage vector-quantized representation $\{q^{(1)}_{1:L_1}, ..., q^{(S)}_{1:L_S}\}$, where $S$ and $L_S$ denote the number of stages and the sequence length at $S$-th stage. It first embeds $x_{1:L}$ to hidden sequences with encoders progressively. Before each encoder block, a strided convolutional layer with the down-sampling rate $d_j$ is applied for down-sampling. H0idden sequences are quantized stagely from the highest stage with the lowest resolution. Firstly, the output sequence from $E_S$ is transformed by a projection layer and quantized to $q^{(S)}_{1:L_S}$ in the quantizer $Q_S$. It is then processed by the residual module $R_S$, and up-sampled by repeating each frame for $d_S$ times. The residual output is concatenated with the next encoder output to calculate the next quantized sequence $q^{(S-1)}_{1:L_{S-1}}$, and added with it using a projection layer to get the next residual output. The remaining sequences are generated in the same way recursively until we obtain $\hat{x}_{1:L}$. Besides, the correlation between two adjacent sequences is enhanced through predicting $q^{(j)}_{1:L_{j}}$ from the higher-stage sequence $q^{(j+1)}_{1:L_{j+1}}$. Hence, the prediction module $P_j$, i.e. a projection layer, is employed to generate the predicted sequence $p^{(j)}_{1:L_{j}}$ from the residual output. It is helpful to train deeper modules effectively \cite{Child2021VeryDV}. Finally, the complete loss function of this model is written as follows:
\begin{equation}
\label{eq:msvq}
\begin{split}
    L_{MS} &= MSE(x, \hat{x})
    + \frac{\alpha}{S}\sum^{S}_{j=1}MSE(h^{(j)}_{1:L_{j}}, sg(q^{(j)}_{1:L_{j}})) \\
    &+ \frac{\beta}{S-1}\sum^{S-1}_{j=1} L_C(p^{(j)}_{1:L_{j}}, sg(q^{(j)}_{1:L_{j}}))
\end{split}
\end{equation}
where $h^{(j)}_{1:L_{j}}$ denotes $j$-th latent sequence before quantization, $L_C$ refers to an arbitrary loss function between two sequences, $\alpha$ and $\beta$ are weight coefficients. 

Finally, MSMCR $\{q^{(j,k)}_{1:L_{j}} | 1 \leq j \leq S, 1 \leq k \leq H\}$ is extracted by multi-stagely modeling speech with multi-head vector quantization, where $q^{(j,k)}_{1:L_{j}}$ denotes the $j$-th quantized sequence at $k$-th head with the length of $L_{j}$.

\section{MSMC-TTS}
\label{sec:tts}

In this section, we will introduce the MSMC-TTS system, including analysis-synthesis, and multi-stage prediction.

\subsection{Analysis-Synthesis}

In analysis, the waveform is first processed to become the acoustic features $x_{1:L}$ via signal analysis, i.e. Mel spectrograms, which has shown respectable completeness in neural vocoders. Then, the feature analyzer is used to process them into MSMCRs. In synthesis, the neural vocoder convert MSMCRs into the corresponding speech waveform. Sequences at different resolutions in MSMCR are up-sampled to the same resolution as $x_{1:L}$ by repetition, and then concatenated together to form the input.

\subsection{Multi-Stage Prediction}

We propose a multi-stage predictor based on FastSpeech \cite{FastSpeech} as the acoustic model of this TTS system.

\subsubsection{Model Architecture}

\begin{figure}[htp]
\centering
\includegraphics[width=8cm]{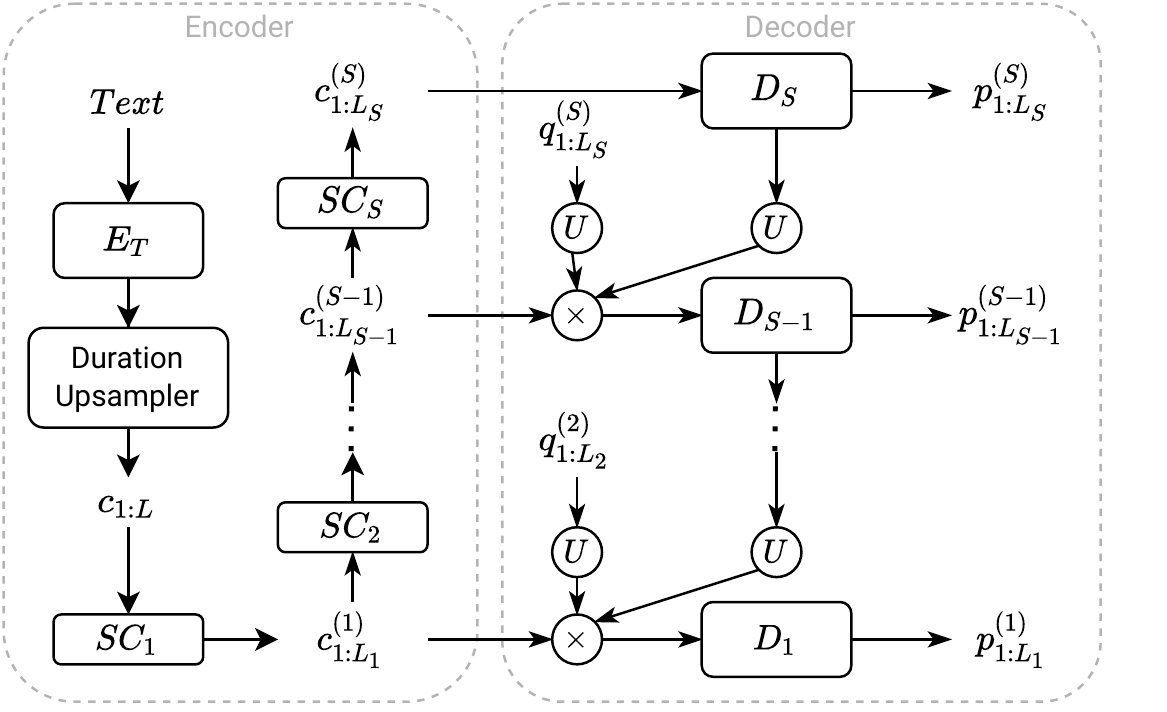}
\caption{The framework of multi-stage predictor. $E_T$ refers to the text encoder. $D_S$, $SC_S$ denote the $S$-th decoder and strided convolutional layer respectively.}
\label{fig:msdec}
\end{figure}

As shown in Fig.\ref{fig:msdec}, the textual features are encoded and up-sampled to $c_{1:L}$ by the duration upsampler, i.e. the length regulator in FastSpeech. $c_{1:L}$ is down-sampled by strided convolutional layers to different resolutions for decoding. First, $c^{(S)}_{1:L_S}$ is fed to the decoder $D_S$ to obtain the first predicted sequence $p^{(S)}_{1:L_S}$. The last hidden output in the decoder is up-sampled by repetition, and concatenated with $q^{(S)}_{1:L_{S-1}}$ and $c^{(S-1)}_{1:L_{S-1}}$ to calculate the next sequence $p^{(S-1)}_{1:L_{S-1}}$. The remaining sequences are generated recursively in the same way. In training, $q^{(j)}_{1:L_j}$ is extracted from the target audio using the analyzer. During inference, we need to replace it with $\hat{q}^{(j)}_{1:L_j}$ quantized from $p^{(j)}_{1:L_j}$ using the corresponding codebooks.

\subsubsection{Loss Function}

To map the input text to the expected prototype vectors in codebooks, the model is trained to estimate them in the continuous space directly. For the predicted vector $p^{(j,k)}_i$, i.e. the $k$-th head of $i$-th vector in the $j$-th sequence, we first adopt MSE loss to make it approach the target $q^{(j,k)}_i$ by reducing the L2 distance: 
\begin{equation}
    L_{MSE} = \frac{1}{S} \sum^S_{j=1} \frac{1}{H} \sum^{H}_{k=1} \frac{1}{L_j} \sum^{L_j}_{i=1}  MSE(p^{(j,k)}_i, q^{(j,k)}_i)
\end{equation}
We also adopt the triplet loss \cite{dong2018triplet}, an effective ranking loss used in metric learning, to ensure that the L2 distance between the input vector and target code is the smallest among all codes in the codebook $C_{j,k}$. The triplet loss is written as:
\begin{equation}
\begin{split}
L_{TPL} &= \frac{1}{S} \sum^S_{j=1} \frac{1}{H} \sum^{H}_{k=1} \frac{1}{L_j} \sum^{L_j}_{i=1}  TPL(p^{(j,k)}_i, q^{(j,k)}_i) \\
TPL&(p^{(j,k)}_i, q^{(j,k)}_i) = \frac{1}{M} \sum^{M}_{ e \ne q^{(j,k)}_i \atop e \in C_{j,k} } max(0, \\
&\left \| p^{(j,k)}_i - q^{(j,k)}_i \right \|_2 - \left \| p^{(j,k)}_i - e \right \|_2 + \epsilon)
\end{split}
\end{equation}
where $\epsilon$ is a constant value called margin number and $e$ refers to a non-target code in the codebook $C_{j,k}$. In this way, the output vector will not only be closer to the target, but also lie farther away from non-target codes. The complete loss function is written as: 
\begin{equation}
    L_{P} = L_{MSE} + \gamma * L_{TPL}
\end{equation}
where $\gamma$ is a weight coefficient.

\section{Experiments}
\label{sec:exp}

This section describes the experiments that evaluate the performance of the proposed TTS system.

\subsection{Experimental Setup}

The standard English single-speaker TTS dataset, Nancy \cite{bc2011}, is used in our experiments. It includes 16 hours of clean recordings of a female voice with corresponding transcripts. The phoneme sequence with punctuations extracted from the text is used as input features. Audios with the sampling rate of 16kHz are converted to 80-dim log-scale Mel spectrograms with a frameshift of 12.5ms and window length of 50ms. Phoneme-level durations are extracted from a pre-trained Tacotron2 model with stepwise attention \cite{he2019robust}.

Neural vocoders in all TTS systems are implemented based on HifiGAN \cite{hifigantts}. UnivNet discriminator \cite{jang2021univnet} is adopted to train the vocoder by both frequency-domain and time-domain discrimination for better audio quality. The MSMC feature analyzer is implemented with FeedForward Transformer blocks \cite{vaswani2017attention}. In each encoder or decoder, the input sequence is processed by a projection layer, and added with position encodings, and then fed to 4 Transformer blocks with the model dimension of 256. The loss function adopts $L_P$ as $L_C$, and sets $\alpha$, $\beta$, $\gamma$ to 1, 0.1, 1. All analyzers are trained for 200,000 iterations using the Adam optimizer \cite{KingmaB14} with the batch size of 64. The learning rate is exponentially decayed from $2 \times 10^{-4}$ to $10^{-6}$ after 20,000 iterations. The acoustic model, multi-stage predictor, is implemented based on FastSpeech, where the dimension of hidden vectors is 600, and the number of Transformer blocks in the text encoder and each decoder is 6. It is trained for 100,000 iterations with the same training configuration as the feature analyzer. Besides, except for the predictor loss $L_P$, the duration loss in FastSpeech \cite{FastSpeech} is also employed to train the duration predictor with a weight coefficient of 0.1.

Finally, mean opinion score (MOS) tests are conducted for subjective evaluation. In each test, 15 listeners are involved to evaluate 20 sets of samples. The listener needs to rate each audio sample according to its overall quality with the score range from 1 to 5 and an increment of 0.5\footnote{Samples are available at \nolinkurl{https://hhguo.github.io/DemoMSMCTTS}}.

\subsection{System Comparison}
\label{ssec:etts}

To evaluate the performance of MSMC-TTS, we compare it with the baseline system, a Mel spectrogram based FastSpeech (Mel-FS) with the same model configuration. For MSMC-TTS, we adopt a two-stage representation with the down-sampling rates $[d_1=1, d_2=4]$, and 4-head vector quantization with the codebook size of 512. \footnote{The implementation of FastSpeech is available at \nolinkurl{ https://github.com/NVIDIA/NeMo}}

\subsubsection{Standard TTS}

We first evaluate their performance in standard single-speaker TTS. Table \ref{tab:mos_sys} shows the test results. The baseline system Mel-FS, obtains the MOS of 3.62, which has an obvious gap compared with the original recordings with the MOS of 4.50. Mel-FS-FT refers to Mel-FS with the vocoder fine-tuned for 100,000 more iterations on the synthetic Mel spectrograms, which are generated from the training set using ground-truth durations. This method is often used to adapt the vocoder to the predicted features. In our experiments, it only improves audio quality slightly to the MOS of 3.69. Compared with the two previous systems, MSMC-TTS shows significantly improved performance with MOS of 4.41, which is very closed to the quality of the original recordings. The generated audio presents higher fidelity and naturalness in different time resolutions, which shows the effectiveness of MSMC-TTS. 

\begin{table}[htp]
\centering
\caption{MOS Test: Standard single-speaker TTS ($\pm$ 95\% CI)}
\label{tab:mos_sys}
\begin{tabular}{cccc}
Recording & Mel-FS & Mel-FS-FT & MSMC-TTS \\ \hline
4.50 $\pm$ 0.08 & 3.62 $\pm$ 0.09 & 3.69 $\pm$ 0.09 & 4.41 $\pm$ 0.11 \\ \hline
\end{tabular}%
\end{table}

\subsubsection{Lightweight TTS}

We evaluate the impact of the decreased modeling complexity on the proposed approach in lightweight TTS by using fewer parameters or simpler model structures. We conduct another MOS test to compare Mel-FS and MSMC-TTS systems with these three model configurations as the encoder or decoder of the acoustic model:
\begin{itemize}
    \item M1: 4 layers of 600-dim Transformer blocks. 
    \item M2: 3 layers of 128-dim Transformer blocks.
    \item M3: 4 layers of 128-dim 1-D convolutional layers.
\end{itemize}

The test results are shown in Table \ref{tab:mos_light}. For Mel-FS, the output quality is degraded seriously from 4.08 to 2.10 when the parameters are reduced from 72.50 MB to 1.75 MB. When Transformer blocks are replaced with the less complex CNNs, the MOS decreases further to 1.86. Both intelligibility and fidelity are affected seriously by the reduced modeling complexity. However, MSMC-TTS still performs well in all situations with only slight degradation as the parameters are reduced. MSMC-TTS trained with M3 (fewest parameters and simplest structure) still  significantly outperforms Mel-FS with M1. It validates that the proposed approach has a much lower requirement to the modeling complexity, and great potential in lightweight TTS.

\begin{table}[htp]
\centering
\caption{MOS Test: Lightweight TTS ($\pm$ 95\% CI)}
\label{tab:mos_light}
\resizebox{0.47\textwidth}{!}{
\begin{tabular}{c|c|ccc}
  \multicolumn{2}{c|}{}  & M1 & M2 & M3 \\ \hline
\multirow{2}{*}{Mel-FS} & Param (MB) & 72.50 & 1.75 & 0.52 \\ \cline{2-5} 
 & MOS & 4.08 $\pm$ 0.10 & 2.10 $\pm$ 0.09 & 1.86 $\pm$ 0.09 \\ \hline
\multirow{2}{*}{MSMC-TTS} & Param (MB) & 116.58 & 4.81 & 3.12 \\ \cline{2-5}
 & MOS & 4.65 $\pm$ 0.08 & 4.58 $\pm$ 0.08 & 4.47 $\pm$ 0.09 \\ \hline
\end{tabular}%
}
\end{table}

\subsection{Ablation Studies}

To verify the effectiveness of multi-head vector quantization and multi-stage representation, MSMC-TTS sytems based on three versions of MSMCRs are compared in terms of completeness, compactness, and TTS quality:
\begin{itemize}
    \item V1: 1-stage representation with 1-head quantization.
    \item V2: 1-stage representation with 4-head quantization.
    \item V3: 2-stage representation with 4-head quantization and down-sample rates of $[1, 4]$.
\end{itemize}
Compactness is represented by the compression ratio over the 80-dim Mel spectrogram. Completeness is evaluated in terms of the MCD, F0-RMSE, and V/UV error rates between the ground-truth audio and the reconstructed audio in analysis-synthesis. Finally, an MOS test is conducted to evaluate their performance in standard single-speaker TTS. 

Table \ref{tab:mos_rep} shows the experimental results. V1, quantized by only one codebook, has the highest compactness with the highest compression ratio of 284. But its completeness is also the worst with the highest MCD of 2.79 dB, and the reconstructed audio cannot even present accurate pronunciations. Hence, its performance in TTS obtains the lowest MOS of 3.74. V2 using multi-head vector quantization has a 4x lower compression ratio of 71, which is still quite high. But the completeness has significantly recovered with much lower MCD, F0-RMSE, and V/UV error rates. Pronunciation and prosody are expressed accurately in reconstruction. The TTS performance is much higher at MOS of 4.36. It shows that improving completeness while keeping high compactness is helpful to raise TTS performance.

Compared with V2, the two-stage representation V3 has slightly reduced compactness at compression ratio of 57. Its completeness is equivalent to V2 with a slightly higher MCD but somewhat reduced loss in F0. However, it achieves the highest MOS of 4.61. It shows obvious improvement on prosody in both coarse-grain and fine-grain levels, which further validates the effectiveness of multi-stage modeling. Hence, both multi-codebooks and multi-stages are crutial for MSMC-TTS.

\begin{table}[htp]
\centering
\caption{Comparsion of different representations on compactness, completeness, and TTS performance.}
\label{tab:mos_rep}
\resizebox{0.45\textwidth}{!}{
\begin{tabular}{c|c|ccc|c}
 & \multirow{2}{*}{\begin{tabular}[c]{@{}c@{}} Compression \\ Ratio \end{tabular}} & \multicolumn{3}{c|}{Analysis-Synthesis} & TTS \\ \cline{3-6}
 &  & \begin{tabular}[c]{@{}c@{}} MCD \\ (dB) \end{tabular} & \begin{tabular}[c]{@{}c@{}} F0-RMSE \\ (Hz) \end{tabular} & \begin{tabular}[c]{@{}c@{}} F0-VUV \\ (\%) \end{tabular} & \begin{tabular}[c]{@{}c@{}} MOS \\ (95\% CI) \end{tabular} \\ \hline
 V1 & 284 & 2.79 & 3.53 & 3.09 & 3.74 $\pm$ 0.09 \\
 V2 & 71  & 2.19 & 2.43 & 2.38 & 4.36 $\pm$ 0.08 \\
 V3 & 57  & 2.22 & 2.42 & 2.33 & 4.61 $\pm$ 0.07 \\ \hline
\end{tabular}%
}
\end{table}

\section{Conclusion}
\label{sec:conclusion}

This paper proposes a MSMC VQ-VAE approach to improve neural TTS. The proposed feature analyzer converts the Mel spectrogram to MSMCRs that are composed of sequences at different time resolutions, and quantized by multiple codebooks, respectively. A multi-stage predictor is trained as the acoustic model to better predict MSMCRs by minimizing the combined loss of MSE and triplet loss. In experiments, MSMC-TTS achieves the highest MOS score of 4.41, which significantly outperforms the baseline system with an MOS of 3.62. Moreover, the proposed approach shows a lower requirement to the modeling complexity, which still performs well with fewer model parameters. Ablation studies validate that multiple stages and multiple codebooks are both crutial for MSMC-TTS.

\section{Acknowledgments}
This work is supported by the Centre for Perceptual and Interactive Intelligence (CPII) Ltd under the Innovation and Technology Fund. Thanks Guanghui Xu for conducting subjective tests.

\bibliographystyle{IEEEtran}
\bibliography{refs}

\vfill

\end{document}